\documentclass[aps,prb,twocolumn,superscriptaddress]{revtex4}
\usepackage{graphicx}

\begin{document}

\title{Electron doping evolution of the anisotropic spin excitations in BaFe$_{2-x}$Ni$_{x}$As$_{2}$}
\author{Huiqian Luo}
\affiliation{Beijing National Laboratory for Condensed Matter
Physics, Institute of Physics, Chinese Academy of Sciences, Beijing
100190, China}
\author{Zahra Yamani}
\affiliation{Canadian Neutron Beam Centre, National Research
Council, Chalk River Laboratories, Chalk River, Ontario K0J 1J0,
Canada}
\author{Yanchao Chen}
\affiliation{Beijing National Laboratory for Condensed Matter
Physics, Institute of Physics, Chinese Academy of Sciences, Beijing
100190, China}
\author{Xingye Lu}
\affiliation{Beijing National Laboratory for Condensed Matter
Physics, Institute of Physics, Chinese Academy of Sciences, Beijing
100190, China}
 \affiliation{ Department of Physics and Astronomy,
The University of Tennessee, Knoxville, Tennessee 37996-1200, USA}
\author{Meng Wang}
\affiliation{Beijing National Laboratory for Condensed Matter
Physics, Institute of Physics, Chinese Academy of Sciences, Beijing
100190, China}
\author{Shiliang Li}
\affiliation{Beijing National Laboratory for Condensed Matter
Physics, Institute of Physics, Chinese Academy of Sciences, Beijing
100190, China}
\author{Thomas A. Maier}
\affiliation{Computer Science and Mathematics Division and Center
for Nanophase Materials Sciences, Oak Ridge National Laboratory, Oak
Ridge, Tennessee 37831 USA}
\author{Sergey Danilkin}
\affiliation{Bragg Institute, Australian Nuclear Science and
Technology Organization, New Illawarra Road, Lucas Heights NSW-2234
Australia}
\author{D. T. Adroja}
\affiliation{ISIS Facility, Rutherford Appleton Laboratory, Chilton,
Didcot, Oxfordshire OX11 0QX, UK}
\author{Pengcheng Dai}
\email{pdai@utk.edu}
\affiliation{ Department of Physics and
Astronomy, The University of Tennessee, Knoxville, Tennessee
37996-1200, USA}
\affiliation{Beijing National Laboratory for
Condensed Matter Physics, Institute of Physics, Chinese Academy of
Sciences, Beijing 100190, China}


\begin{abstract}
We use inelastic neutron scattering to systematically investigate
the Ni-doping evolution of the low-energy spin excitations in
BaFe$_{2-x}$Ni$_{x}$As$_{2}$ spanning from underdoped
antiferromagnet to overdoped superconductor ($0.03\leq x \leq
0.18$).  In the  undoped state, BaFe$_2$As$_2$ changes from
paramagnetic tetragonal phase to orthorhombic antiferromagnetic (AF)
phase below about $138$ K, where the low-energy ($\le\sim80$ meV)
spin waves form transversely elongated ellipses in the $[H, K]$
plane of the reciprocal space.  Upon Ni-doping to suppress the
static AF order and induce superconductivity, the $c$-axis magnetic
exchange coupling is rapidly suppressed and the momentum
distribution of spin excitations in the $[H, K]$ plane is enlarged
in both the transverse and longitudinal directions with respect to
the in-plane AF ordering wave vector of the parent compound. As a
function of increasing Ni-doping $x$, the spin excitation widths
increase linearly but with a larger rate along the transverse
direction. These results are in general agreement with calculations
of dynamic susceptibility based on the random phase approximation
(RPA) in an itinerant electron picture. For samples near optimal
superconductivity at $x\approx 0.1$, a neutron spin resonance
appears in the superconducting state. Upon further increasing the
electron-doping to decrease the superconducting transition
temperature $T_c$, the intensity of the low-energy magnetic
scattering decreases and vanishes concurrently with vanishing
superconductivity in the overdoped side of the superconducting dome.
Comparing with the low-energy spin excitations centered at
commensurate AF positions for underdoped and optimally doped
materials ($x\le 0.1$), spin excitations in the over-doped side
($x=0.15$) form transversely incommensurate spin excitations,
consistent with the RPA calculation.  Therefore, the itinerant
electron approach provides a reasonable description to the
low-energy AF spin excitations in BaFe$_{2-x}$Ni$_{x}$As$_{2}$.
\end{abstract}

\maketitle

\section{Introduction}
Spin excitations are thought to be a candidate for mediating the
electron pairing for superconductivity in unconventional
superconductors \cite{scalapino,Monthoux}. For copper oxide
high-transition temperature (high-$T_c$) superconductors,
superconductivity arises from charge carrier doping of their
antiferromagnetic (AF) Mott insulating parent compounds and forms a
superconducting (SC) dome including underdoped, optimally doped, and
overdoped materials \cite{PALee}.  Although static AF order is
suppressed in optimally doped superconductors, spin excitations
persist throughout the SC dome, and vanish when superconductivity
ceases to exist in the overdoped materials \cite{fujita}. These
results provided compelling evidence that SC electrons in copper
oxides are intimately associated with spin excitations
\cite{jzhao11}, and spin excitations may mediate electron pairing
for superconductivity \cite{kjin}. For iron pnictides
\cite{kamihara,Rotter,assefat,ljli}, superconductivity can also be
induced from the electron or hole doping of their AF ordered
metallic parent compounds \cite{cruz,jzhao,qhuang}.  Because the
parent compounds of iron pnictide superconductors are metallic with
hole and electron Fermi surfaces centered at $\Gamma$ and $M$ points
of the reciprocal space, respectively, the AF order and
superconductivity may arise from quasiparticle excitations between
the hole and electron pockets
\cite{mazin2011n,Hirschfeld,Chubukov,kuroki,fwang}, much different
from the local moment Mott physics of copper oxides \cite{PALee}. In
the electron itinerant picture, the AF order in the parent compounds
arises from Fermi surface nesting of the hole and electron pockets.
Since the electron-doping that induces superconductivity also
increases the size of electron pocket near $M$ points and reduces
the hole-pocket size near $\Gamma$ points, the static AF order is
gradually suppressed with increasing electron doping and
superconductivity emerges from the signed reversed quasiparticles
excitations between the hole and electron pockets
\cite{mazin2011n,Hirschfeld,Chubukov,kuroki,fwang}.  As a
consequence of opening up the electronic gaps at the hole
($\Delta_h$) and electron ($\Delta_e$) Fermi pockets in the SC
state, a neutron spin resonance is expected to occur at the AF
nesting wave vector with an energy $E\le
2\Delta=(\left|\Delta_h\right|+\left|\Delta_e\right|)$
\cite{maier08,maier09,Korshunov}. Indeed, inelastic neutron
scattering experiments on single crystals of electron-doped
BaFe$_{2-x}T_x$As$_2$ ($T=$Co, Ni, see inset in Fig. 1 for the
crystal structure) confirm the presence of the resonance
\cite{mdlumsden,sxchi,sli09,dkpratt09,adChristianson,dsinosov09,mywang10,clester10,jtpark10,hfli10,dsinosov11,mwang11}.
In particular, recent time-of-flight inelastic neutron scattering
measurements reveal that the high-energy ($E>100$ meV) spin waves in
the non-SC BaFe$_2$As$_2$ persist into the SC
BaFe$_{1.9}$Ni$_{0.1}$As$_2$, and the effect of electron-doping is
to form the resonance and modify the low-energy spin excitations
\cite{msliu12}.

\begin{figure}[t]
\includegraphics[scale=.34]{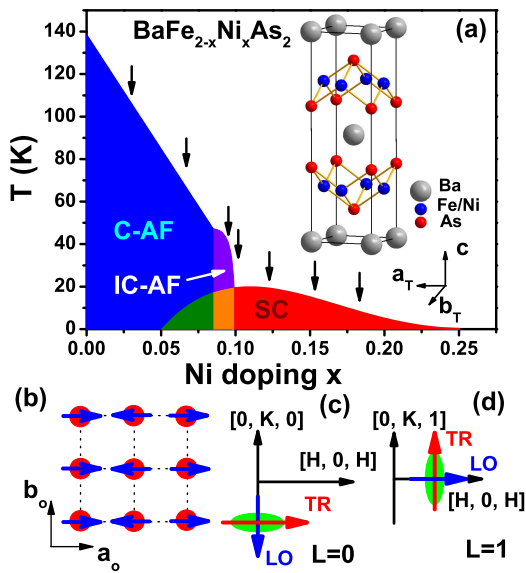}
\caption{(Color online)
(a) The electronic phase diagram  and crystal structure of
BaFe$_{2-x}$Ni$_{x}$As$_{2}$. The arrows indicate seven doping
levels studied in this experiment. (b) the co-linear AF
structure in the FeAs-plane of BaFe$_{2-x}$Ni$_{x}$As$_{2}$. (c) and (d)
Schematic pictures of constant-energy scans along transverse (TR)
and longitudinal (LO) directions both at $L=0$ and $L=1$.
}
\end{figure}

If we assume that the high-energy spin excitations are due to
localized moments and electron doping only affects the Fermi surface
nesting and low-energy spin excitations \cite{msliu12}, one can make
a direct comparison between the measured neutron scattering wave
vector profiles and results from the random phase approximation
(RPA) calculations of the three-dimensional tight-binding model in
the local density approximation (LDA) \cite{graser}.  For example,
the transversely elongated ellipses of the resonance in the
electron-doped BaFe$_{2-x}T_x$As$_2$
\cite{clester10,jtpark10,hfli10,msliu12} from the spin waves in
BaFe$_2$As$_2$ \cite{lharriger} due to the enhancement of the
intraorbital, but interband, pair scattering process between the
$d_{xy}$ orbitals have been interpreted as being more effective in
giving rise to the fully gapped $S^\pm$-symmetry superconductivity
\cite{jzhang10}. The density-functional-theory (DFT) calculations
\cite{jtpark10} had also predicted correctly that the momentum
anisotropy of the neutron spin resonance in the optimally hole-doped
materials is rotated by 90$^\circ$ from that of the electron-doped
case and becomes the longitudinally elongated ellipse
\cite{clzhang}. Moreover, the effect of Fermi surface nesting
appears to account for the hole-doping evolution of the spin
excitations in Ba$_{1-x}$K$_x$Fe$_2$As$_2$ \cite{castellan}.
Finally, recent neutron diffraction work has established, within a
narrow region of $x$ in BaFe$_{2-x}T_x$As$_2$ (Fig. 1a), elastic
transversely incommensurate short-range magnetic peaks that has been
hailed as direct evidence for the spin-density-wave order due to
mismatch of the hole-electron pocket Fermi surface nesting
\cite{pratt11,hqluo,mgkim}.

\begin{figure*}[t]
\includegraphics[scale=.6]{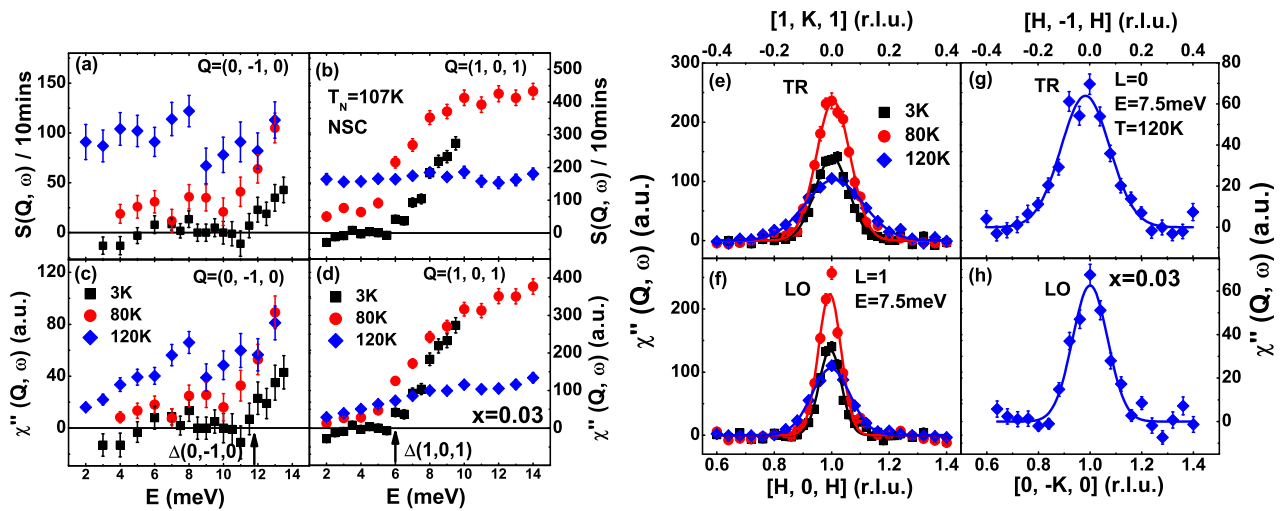}
\caption{(Color online)
Low energy spin excitations of the non-SC
BaFe$_{2-x}$Ni$_{x}$As$_{2}$ ($x=0.03$) with $T_N= 107$ K. (a) and
(b) Energy dependence of $S({\bf Q}, \omega)$ at ${\bf Q}=(0, -1, 0)$ and
${\bf Q}=(1, 0, 1)$ for $T=3, 80, 120$ K, after subtracting the
background at  ${\bf Q}=(0, -1.4, 0)$ and ${\bf Q}=(1.4, 0, 1.4)$, respectively.
(c) and (d) Dynamic spin susceptibility $\chi^{\prime\prime}
({\bf Q}, \omega)$ after considering the Bose factor for $L=0$ and $L=1$.
(e)$\sim$(h) $\chi^{\prime\prime} ({\bf Q}, \omega)$ of constant energy
scans at $T=3, 80, 120$ K and $E= 7.5$ meV. The solid lines are
gaussian fits to the data.
 }
 \end{figure*}

Given that the RPA/DFT calculations have so much success in
describing the elastic magnetic scattering and low-energy spin
excitations in BaFe$_{2-x}T_x$As$_2$, it is surprising that there
are still no quantitative comparison of the doping dependence of the
spin excitation profiles with systematic RPA/DFT calculations.  In
particular, while DFT calculations predicted that spin excitations
in 7.5\% electron-doped  BaFe$_{2-x}T_x$As$_2$ should be
incommensurate along the transverse direction, low-energy spin
excitations seen by neutron scattering show only commensurate
scattering with transversely elongated ellipses
\cite{clester10,jtpark10,hfli10,msliu12}.  Since one can
systematically carry out RPA/DFT calculations to obtain the
imaginary part of the dynamic susceptibility,
$\chi^{\prime\prime}({\bf Q},\omega)$, as a function of
electron-doping $x$ in BaFe$_{2-x}T_x$As$_2$, it is important to
compare the calculation with neutron scattering experiments focusing
on wave vector and energy dependence of the low-energy spin
excitations. In this article, we present a systematic inelastic
neutron scattering and RPA/DFT study on BaFe$_{2-x}$Ni$_{x}$As$_{2}$
covering $x=0.03,0.065,0.092,0.1,0.12,0.15,0.18$ shown as vertical
arrows in the electronic phase diagram of Fig. 1(a) \cite{hqluo}.
Consistent with earlier work
\cite{clester10,jtpark10,hfli10,msliu12}, we find that low energy
spin excitations in BaFe$_{2-x}$Ni$_{x}$As$_{2}$ are anisotropic and
form transversely elongated ellipses at the AF order wave vector.
The peak widths in both the transverse and longitudinal directions
increase linearly with $x$.  For samples near optimal
superconductivity, a neutron spin resonance appears below $T_c$. For
samples at the overdoped side $x=0.15$, the low-energy spin
excitations form two transversely incommensurate peaks. Upon further
increasing electron-doping $x$, the low-energy spin excitations
vanish concurrently with the vanishing superconductivity. We compare
these results with RPA/DFT calculations and find them to be
qualitatively similar. These results indicate an intimate connection
between spin excitations and superconductivity, thus suggesting spin
excitations play an important role for superconductivity in iron
pnictides.

\section{Experiment}

We carried out systematic neutron scattering experiments on
BaFe$_{2-x}$Ni$_x$As$_2$ using C5 thermal neutron triple-axis
spectrometer at Canadian Neutron Beam Center in Chalker River,
Canada. The final neutron energy was set to $E_f=14.56$ meV, with
pyrolytic graphite (PG) as monochromator, analyzer, and filters. The
collimations were set to [none, 0.8$^\circ$, 0.85$^\circ$,
2.4$^\circ$]. High quality single crystals up to centimeter sizes
were grown by FeAs self-flux method \cite{NiNi}, detailed procedure
and sample characterization were published elsewhere \cite{Chen}. We
use the nominal composition to represent the Ni doping level $x$. We
define the wave vector ${\bf Q}$ at ($q_x$, $q_y$, $q_z$) as
$(H,K,L) = (q_xa/2\pi, q_yb/2\pi, q_zc/2\pi)$ reciprocal lattice
units (r.l.u.) using the orthorhombic unit cell, where $a \approx b
\approx 5.62$ \AA, and $c = 12.77$ \AA. In this reciprocal space
notation, the co-linear AF structure of the FeAs layer is shown in
Fig. 1(b), and the in-plane AF Bragg peak position corresponding to
the Fermi surface nesting occurs at $\textbf{Q}=[1, 0]$ or $[0,1]$
r.l.u. due to twinning. Based on previous work \cite{hqluo}, we plot
in Fig. 1(a) the schematic AF order and superconductivity schematic
phase diagram.

For the experiment, we chose seven Ni-doping levels spanning from
the non-SC to over-doped SC samples as marked by the vertical arrows
in Fig. 1(a). These include $x=0.03$ (lightly electron-doped non-SC
sample with $T_N=107$ K) \cite{lharriger2}, 0.065 (underdoped SC
sample with $T_N=72$ K and $T_c=8$ K ), 0.092 (nearly optimal doping
SC sample with static incommensurate short-range order, $T_N=45$ K
and $T_c=19$ K), 0.10 (optimal doping without AF order coexisting
with superconductivity, $T_c=20$ K), 0.12 (overdoped SC sample
without AF order, $T_c=19$ K), 0.15 (overdoped SC sample without AF
order, $T_c=14$ K), 0.18 (heavily overdoped SC sample without AF
order, $T_c=9$ K). In order to properly carry out inelastic neutron
scattering experiments, we prepared 15 to 20 pieces of single
crystals for each Ni-doping levels and coaligned them using E3.  The
total sample mass of each Ni-doping level is about 10-15 grams.
Similar to previous work \cite{jtpark10}, we aligned the samples in
the $[H, 0, H]$ and $[0, -K, 0]$ scattering plane, where the
$c$-axis is about 23.5$^\circ$ from the scattering plane.  In this
geometry, we can probe the wave vector transfers both along the
transverse and longitudinal directions near the in-plane AF
positions ${\bf Q} = (1,0, L)$ or $(0,1,L)$ at $L=0,1$ as shown in
Figs. 1(c) and 1(d).  The co-aligned samples are loaded inside a top
loading close cylce refrigerator for easy exchange of samples.

In order to obtain a complete picture of the doping evolution of the
low-energy spin excitations, we also include some results from our
time-of-flight (TOF) inelastic neutron scattering experiments on the
parent compound BaFe$_2$As$_2$ \cite{lharriger}, optimally doped
$x=0.1$ \cite{msliu12}  and overdoped compounds with $x=0.15$. These
TOF experiments were carried out on the MAPS and MERLIN  TOF chopper
spectrometers at the ISIS facility, Rutherford-Appleton Laboratory,
UK \cite{lharriger,msliu12}.  Part of the data in the $x=0.10$
compound was collected on the TAIPAN thermal neutron triple-axis
spectrometer at the Bragg Institute, Australian Nuclear Science and
Technology Organization. Measurements were done with a fixed
$E_f=14.88$ meV by using PG monochromator, filter and analyzer.

To directly compare the neutron scattering results, we have also
carried out RPA calculations based on a five-orbital tight-binding
model \cite{graser}.  The model is obtained by using
a LDA calculation for
BaFe$_{2-x}$Ni$_x$As$_2$, where the main effect of electron doing
is to shift the chemical potential in a rigid band model.

\begin{figure*}[t]
\includegraphics[scale=0.8]{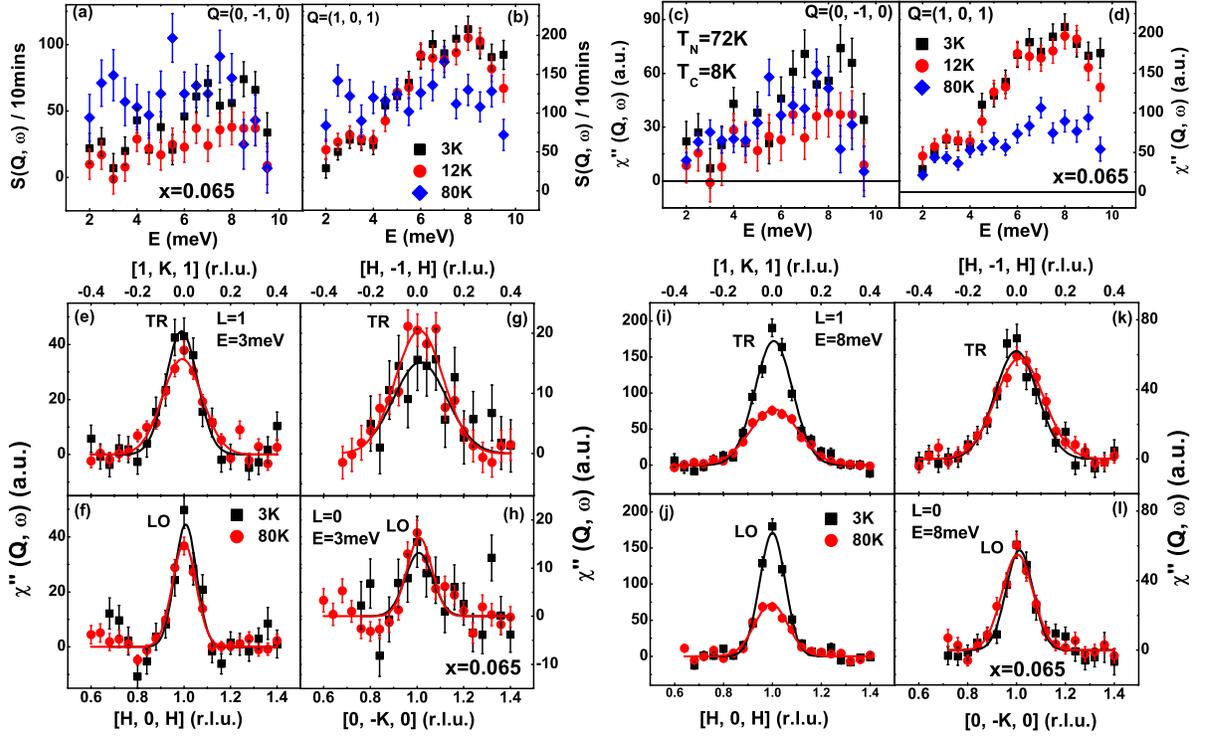}
\caption{(Color online)
Low energy spin excitations of the underdoped
BaFe$_{2-x}$Ni$_{x}$As$_{2}$ ($x=0.065$) with $T_N= 72$ K and $T_c=
8$ K. (a) Energy dependence of $S({\bf Q}, \omega)$ at ${\bf Q}=(0, -1, 0)$ and
(b) ${\bf Q}=(1, 0, 1)$ for $T= 3, 12, 80$ K. (c) and (d)
Corresponding $\chi^{\prime\prime} ({\bf Q}, \omega)$ for $L=0$ and $L=
1$. (e)$\sim$(h) $\chi^{\prime\prime} (Q, \omega)$ of constant
energy scans at $E= 3$ meV and (i)$\sim$(l) $E = 8$ meV for $T= 3$ K
and 80 K.
}
\end{figure*}

\section{Results}

We first describe the inelastic neutron scattering results on
BaFe$_{2-x}$Ni$_x$As$_2$ for $x=0.03$. In previous work on the
$x=0.04$ with $T_N=91$ K and filamentary superconductivity, the
effect of electron-doping is found to reduce the $c$-axis exchange
coupling in BaFe$_2$As$_2$ and induce quasi-two-dimensional spin
excitations \cite{lharriger2}. The anisotropy spin gaps at wave
vectors ${\bf Q}= (1,0,1)$ and $(1,0,0)$ were 2 and 4 meV,
respectively \cite{lharriger2}. For comparison, the AF N${\rm
\acute{e}}$el temperature of the system changes from  $T_N=138$ K
for BaFe$_2$As$_2$ to $T_N=107$ K for
BaFe$_{1.97}$Ni$_{0.03}$As$_2$.  In addition, there are no evidence
for bulk superconductivity until $x=0.05$. From recent inelastic
neutron scattering work on BaFe$_2$As$_2$, the spin-wave gaps at the
Brillouin zone center and boundary are found to be near
$\Delta(1,0,1)=10$ and $\Delta(1,0,0)=20$ meV, respectively
\cite{jtpark12}. Assuming that the nearest neighbors, next nearest
neighbor, and $c$-axis magnetic exchange couplings are $J_{1a}$
($J_{1b}$), $J_2$, and $J_c$, respectively \cite{lharriger,jzhao09},
one can fit spin waves of BaFe$_2$As$_2$ using the Heisenberg
Hamiltonian \cite{lharriger}.   The anisotropic spin gaps are
related to the exchange couplings via \cite{lharriger2} $\Delta(1,
0, 1)=2S[(J_{1a}+2J_2+J_c+J_s)^2-(J_{1a}+2J_2+J_c)^2]^{1/2}$,
$\Delta (1, 0, 0)= \Delta (0, -1,
0)=2S[(2J_{1a}+4J_2+J_s)(2J_c+J_s)]^{1/2}$, where $J_s$ is the
magnetic single ion anisotropy and $S$ is the magnetic spin ($S=1$).
For BaFe$_2$As$_2$, the exchange couplings are estimated from global
fitting of the in-plane high-energy spin waves using the Heisenberg
Hamiltonian \cite{lharriger}.  While the best fitting exchange
energies are $SJ_{1a}= 59.2$ meV,  $SJ_{1b}= -9.2$ meV, $SJ_2= 13.6$
meV, $SJ_c= 1.8$ meV and $SJ_s$=0.084 meV, these results are
obtained from high-energy in-plane spin wave dispersions and
therefore cannot accurately determine the $c$-axis exchange energy.
Recently, using the in-plane magnetic exchange couplings determined
from the high-energy spin wave data \cite{lharriger} and more
accurate measurements of $\Delta(1, 0, 1)$ and $\Delta(1, 0, 0)$,
the effective $c$-axis exchange energy and the single ion anisotropy
of BaFe$_2$As$_2$ are found to be $SJ_c=0.22$ meV and $SJ_s=0.14$
meV, respectively \cite{jtpark12}.  These results are consistent
with our earlier estimations \cite{lharriger2}.

Figures 2(a) and 2(b) show the background subtracted energy scans
$S({\bf Q},\omega)$ at the AF zone boundary ${\bf Q} =(0, -1,
0)\approx (1,0,0)$ and zone center ${\bf Q} =(1, 0, 1)$ for
BaFe$_{1.97}$Ni$_{0.03}$As$_2$, where the background scattering was
measured at ${\bf Q} =(0, -1.4, 0)$ and ${\bf Q} =(1.4, 0, 1.4)$,
respectively. The low-temperature ($T=3$ K) spin-wave gaps at the
zone center and boundary decrease to $\Delta(1,0,1)=5.5$ meV and
$\Delta(0,-1,0)=11$ meV, respectively [Figs. 2(a) and 2(b)]. Upon
warming up to $T=80\ {\rm K}\approx 0.75T_N$, the spin-gap values
decrease rapidly for $\Delta(1,0,1)$ but remains large for
$\Delta(0,-1,0)$. They completely vanish at $T=120\ {\rm K}\approx
1.12T_N$ [Fig. 2(a)]. Figures 2(c) and 2(d) show the energy
dependence of the imaginary part of the dynamic susceptibility,
$\chi^{\prime\prime}({\bf Q},\omega)$, estimated via
$\chi^{\prime\prime}({\bf
Q},\omega)=[1-\exp(-\hbar\omega/k_BT)]S({\bf Q},\omega)$, where
$k_B$ is the Boltzmann constant. While the $\chi^{\prime\prime}({\bf
Q},\omega)$ shows clear spin-wave gaps at 3 K, it increases linearly
with energy in the paramagnetic state at $T=120$ K.

\begin{figure*}[t]
\includegraphics[scale=.6]{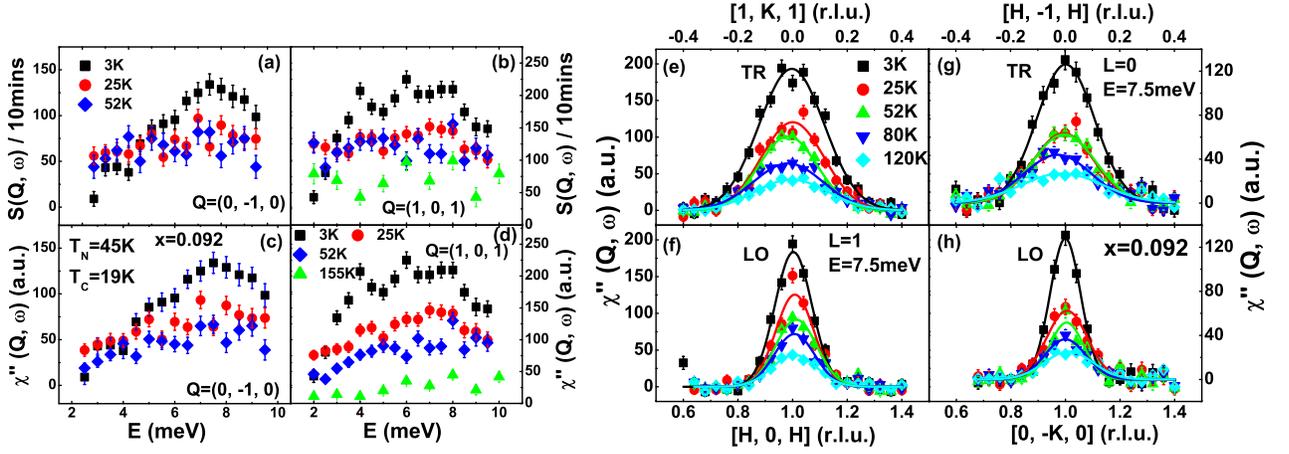}
\caption{
(Color online)
Low energy spin excitations of the nearly optimal doped
BaFe$_{2-x}$Ni$_{x}$As$_{2}$ ($x=0.092$) with $T_N= 45$ K and $T_c=
19$ K. (a) Energy dependence of $S({\bf Q}, \omega)$ at ${\bf Q}=(0, -1, 0)$ and
(b) ${\bf Q}=(1, 0, 1)$ for $T=3, 25, 52$ K and 155 K. (c) and (d)
Corresponding $\chi^{\prime\prime} ({\bf Q}, \omega)$ for $L=0$ and $L=1$.
(e) $\chi^{\prime\prime} ({\bf Q}, \omega)$ of constant energy scans at
$T=3, 25, 52, 80, 120$ K and $E= 7.5$ meV.
}
\end{figure*}

From the recent work on spin excitations of optimally electron-doped
BaFe$_{1.9}$Ni$_{0.1}$As$_{2}$ superconductor \cite{msliu12}, we see
that the high-energy spin excitations, and therefore the effective
in-plane magnetic exchange energies, are not affected by
electron-doping and superconductivity. Assuming that the in-plane
magnetic exchange couplings in BaFe$_{1.97}$Ni$_{0.03}$As$_2$ are
unchanged from those of the BaFe$_2$As$_2$ \cite{lharriger}, we
estimate $SJ_c= 0.066$ meV and $SJ_s\approx 0$ meV using the newly
measured $\Delta(1,0,1)$ and $\Delta(1,0,0)$ values for
BaFe$_{1.97}$Ni$_{0.03}$As$_2$ [Figs. 2(a) and 2(b)]. These results
are consistent with the notion that electron-doping in
BaFe$_{2-x}$Ni$_x$As$_2$
 rapidly decreases the out-of-plane exchange couplings
\cite{lharriger2}.

\begin{figure}[t]
\includegraphics[scale=.6]{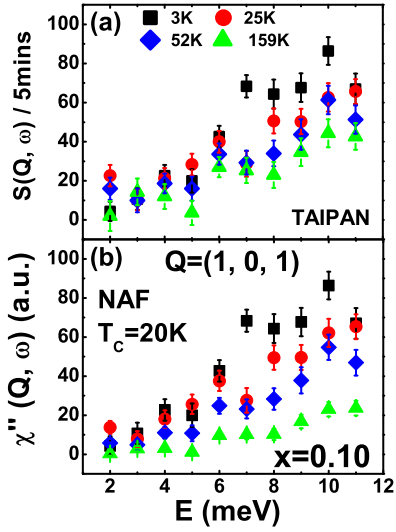}
\caption{
(Color online)
(a) and (b) Energy scans and corresponding
$\chi^{\prime\prime} ({\bf Q}, \omega)$ at $Q=(1, 0, 1)$ and $T=3, 25,
52, 82, 159$ K for the optimally doped BaFe$_{2-x}$Ni$_{x}$As$_{2}$
($x=0.10$) with $T_c= 20$ K. These data are collected at TAIPAN
triple-axis spectrometer in ANSTO.
 }
\end{figure}

To explore the in-plane momentum distribution of the spin
excitations in BaFe$_{1.97}$Ni$_{0.03}$As$_2$, we carried out
constant-energy scans along the longitudinal $[H,0,H]$ and
transverse $[0,-K,0]$ directions at $E=7.5$ meV and various
temperatures. Figures 2(e)-2(h) show $\chi^{\prime\prime}({\bf
Q},\omega)$ along different directions at $T= 3$, 80, and 120 K.
Around the AF Bragg wave vector ${\bf Q}=(1,0,1)$, spin waves are
weakly anisotropic, being broader along the transverse direction
than that of the longitudinal direction [Figs. 2(e) and 2(f)]. The
scattering intensity increases on warming to 80 K due to closing of
the spin anisotropy gap.  As expected, the spin excitations peak
widths above $T_N$ are broader than the widths of spin waves in the
AF ordered state.  For $L=0$, we cannot find any magnetic scattering
at $E=7.5$ meV below $T_N$ due to the presence of the spin gap.
Transversely elongated paramagnetic scattering emerges at
temperatures above $T_N$ [Figs. 2(g) and 2(h)].

Figure 3 summarizes the constant-energy and constant-${\bf Q}$ scans
for the electron underdoped BaFe$_{1.935}$Ni$_{0.065}$As$_{2}$ with
$T_N=72$ K and $T_c=8$ K \cite{hqluo,Chen}. Figures 3(a) and 3(b)
show the background subtracted magnetic scattering in terms of
energy scans at the AF wave vectors ${\bf Q}=(0,-1,0)$ and ${\bf
Q}=(1,0,1)$ at various temperatures. Although resistivity and
diamagnetic measurements show a SC phase transition below $T_c$= 8 K
in this compound \cite{Chen}, magnetic scattering $S({\bf
Q},\omega)$ at ${\bf Q}=(1,0,1)$ shows no temperature dependence
across $T_c$ (between 3 K and 12 K) in the energy range 2 meV $<E<$
10 meV, thus indicating no neutron spin resonance. Upon warming up
to 80 K ($=T_N+8$ K), the paramagnetic spin excitations at ${\bf
Q}=(0,-1,0)$ and ${\bf Q}=(1,0,1)$ become $L$-independent, in
contrast to the larger magnetic scattering intensity at ${\bf
Q}=(1,0,1)$ in the AF ordered state. Figures 3(c) and 3(d) show the
corresponding $\chi^{\prime\prime}({\bf Q},\omega)$ at ${\bf
Q}=(0,-1,0)$ and ${\bf Q}=(1,0,1)$, respectively.  In contrast to
$\chi^{\prime\prime}({\bf Q},\omega)$ for
BaFe$_{1.97}$Ni$_{0.03}$As$_2$, the spin excitations are gapless at
all temperatures and increase approximately linearly with increasing
energy.

Figures 3(e), 3(f), 3(g), and 3(h) show constant-energy scans below
$T_c$ (3 K) and above $T_N$ (80 K) along the longitudinal and
transverse directions at $E=3$ meV and $L=0,1$.  At $T=3$ K, there
are more magnetic scattering centered around ${\bf Q} = (1,0,1)$
than at  ${\bf Q} = (0,-1,0)$, suggesting the presence of intensity
modulation along the $c$-axis. On warming to 80 K above $T_N$,
$\chi^{\prime\prime}({\bf Q},\omega)$ becomes similar at these two
wave vectors.   Figures 3(i), 3(j), 3(l), and 3(k) plot identical
scans as those of Figs. 3(e)-3(h) at $E=8$ meV. Similar to data at
$E=3$ meV, we see that the large differences in
$\chi^{\prime\prime}({\bf Q},\omega)$ at $L=1$ and 0 at 3 K
essentially vanish on warming up to 80 K.  These results suggest a
weak $L$ modulation of the paramagnetic scattering compared with
spin waves below $T_N$. Comparing ${\bf Q}$-scan data at $E=3$ and 8
meV, we see that the widths of transverse scans increase with
increasing energy, consistent with earlier results on spin waves of
BaFe$_2$As$_2$ \cite{lharriger}.

We now examine spin excitations of BaFe$_{2-x}$Ni$_{x}$As$_{2}$ in a
narrow regime where the transverse incommensurate AF order was found
\cite{hqluo}. For this purpose, we choose
BaFe$_{1.908}$Ni$_{0.092}$As$_{2}$ which has $T_c= 19$ K and
$T_N=40$ K  \cite{hqluo}. Figures 4(a)-4(d) show energy dependence
of the magnetic scattering $S({\bf Q},\omega)$ and
$\chi^{\prime\prime}({\bf Q},\omega)$ at ${\bf Q}=(0,-1,0)$ and
${\bf Q}=(1,0,1)$ below and above $T_c$. While
$\chi^{\prime\prime}({\bf Q},\omega)$ at ${\bf Q}=(0,-1,0)$ appears
to increase approximately linearly with increasing energy at 25 K
and 52 K, the effect of superconductivity is to suppress low-energy
spin excitations and induce a neutron spin resonance near $E=7.5$
meV [Fig. 4(c)].  At ${\bf Q}=(1,0,1)$, $\chi^{\prime\prime}({\bf
Q},\omega)$ behaves similarly except that the superconductivity
induced resonance is rather broad in energy extending from 3 meV to
9 meV, giving a resonance energy of $E\approx 6$ meV [Fig. 4(d)].
These results are consistent with the earlier work \cite{mywang10}.

Since elastic scattering in BaFe$_{1.908}$Ni$_{0.092}$As$_{2}$ forms
static short-range transverse incommensurate AF order \cite{hqluo},
it is interesting to see if one can also find incommensurate spin
excitations. Figures 4(e)-4(h) summarize the transverse and
longitudinal scans across the in-plane AF wave vector at $L=1$ and 0
near the resonance energy of $E=7.5$ meV.  The intensity gain of the
neutron spin resonance below $T_c$ is seen in both the transverse
[Figs. 4(e) and 4(g)] and longitudinal scans [Figs. 4(f) and 4(h)].
Although the widths of transverse scans are much broader than that
of the longitudinal scans, there is no evidence for incommensurate
spin excitations at the resonance energy.  On warming to higher
temperatures, we see gradual reduction of the magnetic scattering
and there are still no evidence for incommensurate spin excitations.
In addition, the ${\bf Q}$-widths of spin excitations are weakly
temperature dependent below 120 K.  Therefore, one can safely assume
that the static short-range incommensurate AF order below $T_N=40$ K
has little impact to spin excitations of
BaFe$_{1.908}$Ni$_{0.092}$As$_{2}$.

For optimally electron-doped BaFe$_{1.9}$Ni$_{0.1}$As$_{2}$
($T_c=20$ K) \cite{sxchi},since recent neutron TOF experiments
\cite{msliu12} have already mapped out the wave vector and energy
dependence of spin excitations throughout the Brillouin zone, we
will not repeat them here but instead focusing on temperature
dependence of the energy scans at ${\bf Q}=(1,0,1)$. Figures 5(a)
and 5(b) show the magnetic scattering $S({\bf Q},\omega)$ and
$\chi^{\prime\prime}({\bf Q},\omega)$ at $T=3$, 25, 52, and 159 K,
respectively.  While one can clearly see the presence of a spin gap
and a resonance near $E\approx 7$ meV at 3 K, the normal state
$\chi^{\prime\prime}({\bf Q},\omega)$ is gapless and increases
linearly with increasing energy.  The  $\chi^{\prime\prime}({\bf
Q},\omega)$ also decreases monotonically with increasing
temperature.

Figure 6 summarizes the transverse and longitudinal scans around the
resonance energy ($E=7.5$ meV) for a slightly electron overdoped
sample, BaFe$_{1.88}$Ni$_{0.12}$As$_{2}$ ($T_c=19$ K).  This sample
has the same SC transition temperature as that of
BaFe$_{1.908}$Ni$_{0.092}$As$_{2}$ but without static AF order.
Similar to constant-energy scans in the underdoped samples, we again
carried out transverse and longitudinal scans along the $[1,K,1]$,
$[H,-1,H]$ and $[H,0,H]$, $[0,-K,0]$ directions, respectively, below
and above $T_c$. Inspection of Figure 6 immediately reveals the
magnetic intensity gain at wave vectors ${\bf Q}=(1,0,1)$ and ${\bf
Q}=(0,-1,0)$ below $T_c$.  We also note that the widths of
transverse scans are considerably broader than that of the
longitudinal scans.  Again, there is no evidence for incommensurate
spin excitations at the resonance energy to within our instrumental
resolution.

\begin{figure}[t]
\includegraphics[scale=.6]{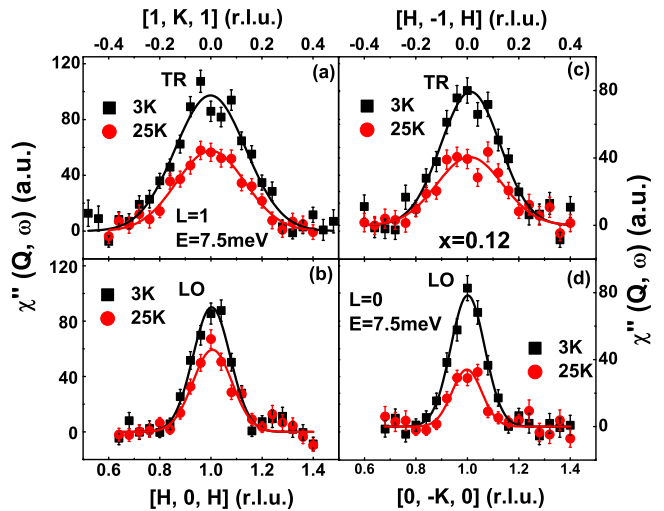}
\caption{ (Color online)
$\chi^{\prime\prime} ({\bf Q}, \omega)$ of constant energy scans
at $T=3, 25$ K and $E= 7.5$ meV for the slightly overdoped
BaFe$_{2-x}$Ni$_{x}$As$_{2}$ ($x=0.12$) with $T_c= 19$ K.
}
\end{figure}

Turning our attention to a more electron overdoped sample
BaFe$_{1.85}$Ni$_{0.15}$As$_{2}$ with $T_c=14$ K, we note that in
previous polarized neutron scattering experiment on these samples
\cite{mliu12}, spin excitations are found to be paramagnetic and
isotropic in space below and above $T_c$.  In the normal state (20
K), the $\chi^{\prime\prime}({\bf Q},\omega)$ at ${\bf Q}=(1,0,1)$
is gapless and increases linearly with energy.  Upon entering into
the SC state (3 K), the $\chi^{\prime\prime}({\bf Q},\omega)$ has a
small spin gap of $\sim$2 meV and a neutron spin resonance at
$E\approx 7$ meV \cite{mliu12}.  Similarly, the
$\chi^{\prime\prime}({\bf Q},\omega)$ at ${\bf Q}=(1,0,2)$ has a
spin gap of $\sim$2 meV and increases with increasing energy
\cite{mliu12}.  Since previous polarized neutron scattering
experiments have already measured the energy dependence of the
$\chi^{\prime\prime}({\bf Q},\omega)$ at AF wave vectors, we will
not repeat them here but instead focus on the in-plane wave vector
anisotropy of the resonance.  Figures 7(a) and 7(b) show the
transverse and longitudinal scans along the $[1,K,1]$ and $[H,0,H]$
directions below and above $T_c$ at the resonance energy of $E=7$
meV.  While one can see an enhancement of magnetic scattering below
$T_c$ due to the resonance, the low-temperature (3 K) transverse
scan also show a flat top consistent of having two incommensurate
peaks instead of one Gaussian.  Transverse scan along the $[H,-1,H]$
direction confirm this conclusion and show two clear incommensurate
peaks at $(-\delta,-1,-\delta)$ and  $(\delta,-1,\delta)$ with
$\delta=0.098$ [Fig. 7(c)].  On the other hand, longitudinal scans
show commensurate peaks centered at the AF wave vector [Figs. 7(b)
and 7(d)].

\begin{figure}[t]
\includegraphics[scale=.5]{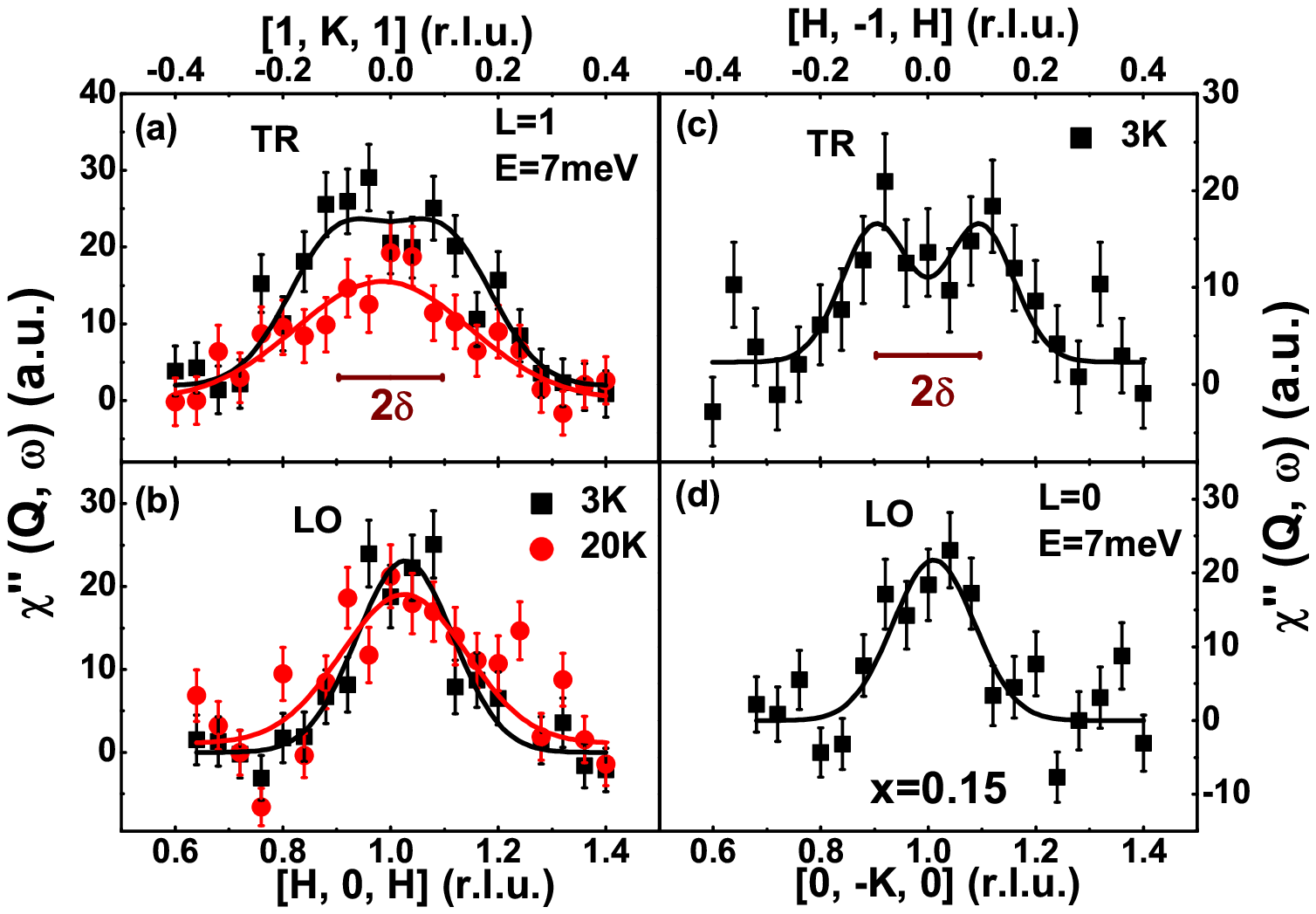}
\caption{ (Color online)
Low energy spin excitations of the overdoped
BaFe$_{2-x}$Ni$_{x}$As$_{2}$ ($x=0.15$) with $T_c= 14$ K.  (a)
$\sim$ (d) Constant energy scans at $T=3, 20$ K and $E= 7$ meV,
where the transverse scans at $T=3$ K show a small
incommensurability with $\delta = 0.098$.
}
\end{figure}

To further probe the possible incommensurate magnetic excitations,we
carried out additional transverse scans at different energies and
compare these results with cuts from the neutron TOF measurements on
the same sample. Figures 8(a) and 8(b) show transverse scans along
the $[1,K,1]$ direction at $E=5$ and 9 meV, respectively. In the
low-temperature SC state, there are two clear incommensurate peaks
as shown by the solid line fits using two Gaussians.  From these
Gaussian fits to the data, we can obtain energy dependence of the
incommensurability, giving $\delta=0.073$ for $E=5$ meV and
$\delta=0.121$ for $E=9$ meV at 3 K. In addition, the incommensurate
peaks appear to be more robust in the SC state. To confirm such a
conclusion, we used neutron TOF chopper spectrometer MERLIN at ISIS,
Rutherford-Appleton Laboratory, to measure spin excitations in
BaFe$_{1.85}$Ni$_{0.15}$As$_{2}$.  The incident beam energy was set
to $E_i=25$ meV, and the sample was aligned such that the angles
between ${\bf k}_i$ and $c$-axis are $\theta=7^\circ, 11^\circ,
16^\circ$.  The two-dimensional wave vector dependent profile of the
spin excitations for $6\ {\rm meV}<E<8\ {\rm meV}$ at $L=1$ can be
covered in this arrangement. The wave vector cuts at $E=7$ meV
obtained by Horace software using the combination of three sets of
TOF data are shown in Figs. 8(c) and 8(d).  In the SC state, the
transverse scan has a flattish top consistent with the triple-axis
data in Fig. 7(a).  On warming to 20 K ($T=6+T_c$), the scattering
shows a broad Gaussian centered around the AF ordering wave vector
[Fig. 8(d)].   Although these results suggest that spin excitations
of BaFe$_{1.85}$Ni$_{0.15}$As$_{2}$ change from the commensurate in
the normal state to incommensurate in the SC state, it remains
unclear if the commensurate-to-incommensurate transition occurs at
$T_c$.

\begin{figure}[t]
\includegraphics[scale=.33]{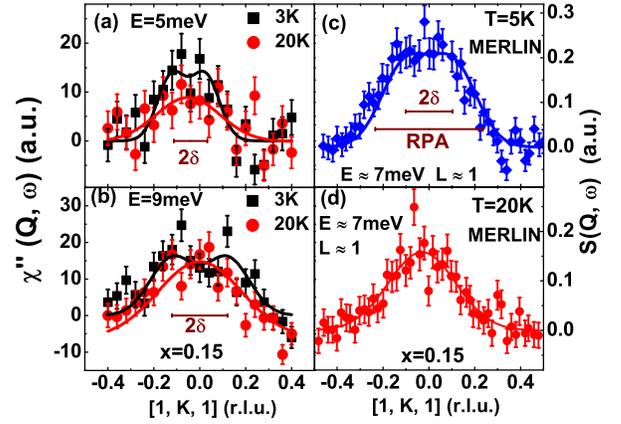}
\caption{ (Color online)
(a) and (b) Transverse ${\bf Q}$ scans at $L=1$, $E= 5$ meV and
9 meV below $T_c$ and above $T_c$. (c) and (d) are constant energy
cuts from the TOF data on MERLIN with energy range of 6
meV$<E<$8 meV and $0.7<L<1.3$ at $T=5$ K and 20 K. The center bar
show the incommensurability $\delta$= 0.073, 0.121 for $E= 5$ meV
and 9 meV, $\delta$= 0.102 for the TOF cut around 7 meV, and $\delta$=
0.235 for RPA calculation, respectively.
}
\end{figure}

\begin{figure*}[t]
\includegraphics[scale=.5]{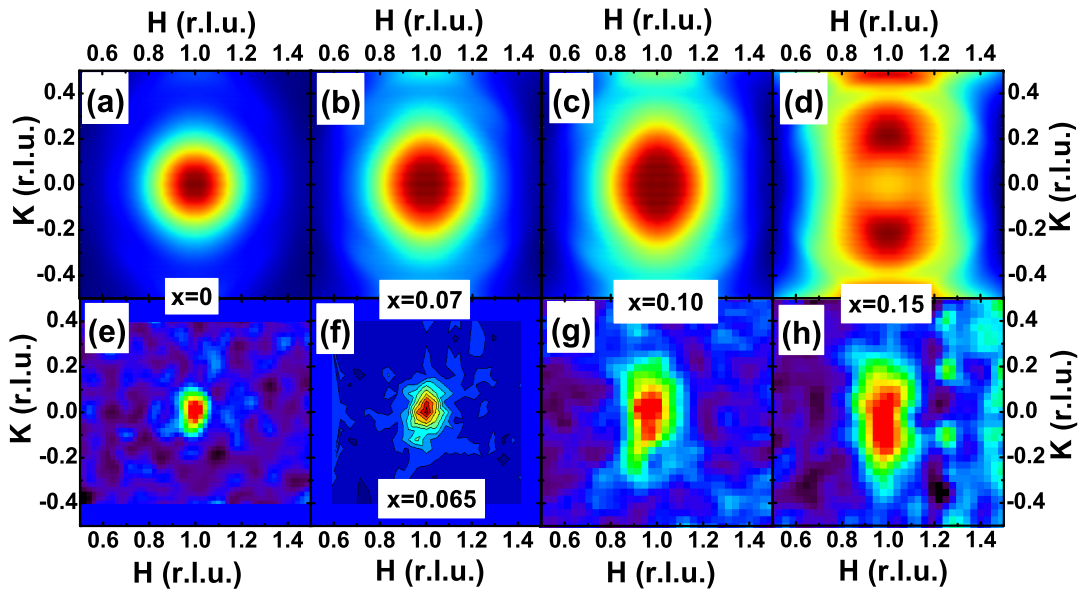}
\caption{ (Color online)
Comparison between the RPA calculation (a)$\sim$(d) and
experimental results (e)$\sim$(h) of the in-plane anisotropic spin
excitations in BaFe$_{2-x}$Ni$_x$As$_2$.
 }
\end{figure*}

Finally, we have searched for spin excitations in heavily overdoped
SC BaFe$_{1.82}$Ni$_{0.18}$As$_{2}$, which has no AF order and
$T_c=9$ K. For the experiment, we have co-aligned about 8 grams of
single crystals.  In spite of much efforts, we have been unable to
find large enough magnetic scattering near the AF wave vector for
energies around 8 meV.  Although this does not mean that there are
no magnetic scattering at this energy, it does suggest a dramatic
reduction in magnetic scattering with increasing Ni-doping in the
overdoped regime.

Using a RPA for a three-dimensional 5-orbital tight-binding model
for BaFe$_2$As$_2$, we have carried out calculations of the RPA spin
susceptibility $\chi^{\prime\prime}({\bf Q},\omega)$ for the normal
state. This model was introduced by Graser {\it et al.}\cite{graser}
and obtained from fits of the DFT band structure for BaFe$_2$As$_2$.
The RPA spin susceptibility $\chi^{\prime\prime}({\bf Q},\omega)$ is
obtained from the RPA multi-orbital susceptibility matrix $\chi^{\rm
RPA}_{l1,l2,l3,l4}$ for orbitals $l_1, l_2, l_3$ and $l_4$, which is
related to the bare (Lindhard) susceptibility matrix
$\chi^0_{l1,l2,l3,l4}$ and the Coulomb interaction matrix $U$
through $\chi^{\rm RPA}({\bf Q},\omega) = \chi^0({\bf
Q},\omega)[1-U\chi^0({\bf Q},\omega)]^{-1}$.  The interaction matrix
$U$ in orbital space is defined in Graser {\it et al.} \cite{graser}
and contains on-site matrix-elements for the intra- and
inter-orbital Coulomb repulsions $U$ and $U^\prime$, and for the
Hunds-rule coupling and pair-hopping terms $J$ and $J^\prime$. For
this calculation, we have used spin-rotationally invariant
parameters $U^\prime=U-2J$ and $J^\prime=J$ with $U = 0.8$ eV and $J
= 0.2$ eV. The effect of Ni substitution in BaFe$_{2-x}$Ni$_x$As$_2$
is assumed to be electron-doping via the rigid band shift.

The top row in Fig. 9 shows results of this calculation obtained for
an energy $E=8$ meV for different electron dopings. As the doping
increases from (a) to (d), one clearly sees an enhancement of the
anisotropy in spin excitations (transverse elongation),
qualitatively similar to the experimental results. For a doping of
$x=0.15$, we find two transverse incommensurate peaks near the AF
wave vector. The corresponding TOF and triple-axis inelastic neutron
scattering measurements at $E=8$ meV are shown in Figs. 9(e)-9(h)
\cite{lharriger,msliu12}. The data for $x=0.065$ were mapped out on
the C5 triple-axis spectrometer with $E=8$ meV and $L=1$ at $T=3$ K
[Fig. 9(f)]. Figure 9(e) is obtained by doing ${\bf Q}$-cut of
energy range 7 meV $\le E\le$ 11 meV from the $x=0$ compound
measured on MAPS at $T=7$ K \cite{lharriger}. Figures 9(g) and 9(h)
are ${\bf Q}$-cuts of the energy range 6 meV $\le E\le$ 9 meV for
the optimal doped compound $x= 0.1$ \cite{msliu12} and the overdoped
compound $x= 0.15$ measured on MERLIN at $T= 5$ K, respectively.
Spin excitations form transversely elongated ellipses that increase
with increasing electron-doping.

\begin{figure}[t]
\includegraphics[scale=.45]{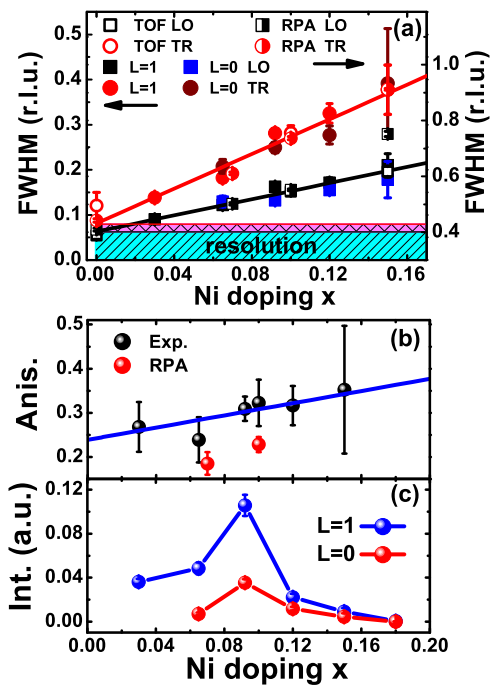}
\caption{
(Color online)
(a) Linear doping dependence of FWHM for the transverse and
longitudinal wave vector scans. (b) Doping dependence of the
in-plane anisotropy of spin excitations. (c) Doping dependence of
peak intensity at ${\bf Q}=(0, -1, 0)$ and (1, 0, 1) normalized by phonon
intensity at $Q=(2, 0.3, 2)$ and $T= 80$ K. The lines are
guides to the eyes.
}
\end{figure}

On initial inspection, it appears that the RPA calculated
$\chi^{\prime\prime}({\bf Q},\omega)$ spectra in Figs. 9(a)-9(d)
occupy much larger portion of the Brillouin zone than that of the
experimental results shown in Figs. 9(e)-9(h). In addition, the RPA
calculated $\chi^{\prime\prime}({\bf Q},\omega)$ in full width half
maximum (FHWM) and incommensurability $\delta$ are about a factor of
2 larger than that of the measurements. However, when the Ni-doping
dependence of the FWHM from the fits to transverse and longitudinal
scans was plotted in Fig. 10(a), we see a well-defined linear
dependence of FWHM versus $x$ for both directions at $L=0$ and
$L=1$.  These results are also consistent with the FHWM deduced from
the TOF measurements discussed in Fig. 9. Although the RPA
calculation gives the absolute values of FWHM that are about a
factor of two larger than that of the experiments, the calculated
spin excitation width also has a linear Ni-doping dependence
consistent with the experiments. To estimate the electron-doping
dependence of the intrinsic spin excitation {\bf Q}-widths at $E=8$
meV, we assume that spin waves in the parent compound BaFe$_2$As$_2$
are instrumental resolution limited at $E=8$ meV
 \cite{lharriger}.  Guassian fits to the scattering profiles along the longitudinal and transverse directions at $E=8$ meV for
BaFe$_2$As$_2$ give $R_{LO}=0.0631\pm 0.0037$ and $R_{TR}=0.0811\pm
0.0053$ r.l.u., respectively. These values are consistent with the
calculated instrumental resolution and previous results on
BaFe$_{2-x}$Co$_x$As$_2$\cite{jtpark10}.

Figure 10(a) shows the Ni-doping dependence of the longitudinal and
transverse spin excitation widths at $E=8$ meV. It is clear that the
excitation widths increase linearly with increasing electron-doping
along both directions, but with a smaller slope along the
longitudinal direction.  This is consistent with the electron-doping
dependence of the $\chi^{\prime\prime}({\bf Q},\omega)$ from the RPA
calculation (solid lines). Figure 10(b) plots the electron doping
dependence of the spin excitation anisotropy ratio $A$, defined as
$A=(W_{TR}-W_{LO})/(W_{TR}+W_{LO})$ where $W_{LO}$ and $W_{TR}$ are
intrinsic widths of spin excitations along the longitudinal and
transverse directions, respectively.
 When the Ni-doping increases from $x=0.03$ to 0.15,
the measured spin excitation anisotropy increases slightly from 0.25
to 0.35. This is consistent with the RPA calculation, thus
suggesting that the RPA calculation captures the essential physics
in the doping dependence of the low-energy spin excitations. Figure
10(c) shows the Ni-doping dependence of the low-temperature
integrated magnetic scattering near $E=8$ meV, where we have
normalized the scattering intensity for different doping levels via
phonons at ${\bf Q}=(2, 0.3, 2)$ and 80 K, at ${\bf Q}=(1,0,L)$ with
$L=0,1$. We see that the magnetic scattering for over-doped samples
decrease systematically with increasing doping, and appear to vanish
around $x= 0.20$ near the overdoped border of superconductivity dome
[Fig. 1(a)].  This is consistent with previous nuclear magnetic
resonance measurements on BaFe$_{2-x}$Co$_x$As$_2$, where the
presence of the enhanced AF spin excitations appears to be
intimately associated with the dome of superconductivity, and vanish
for overdoped sample without superconductivity \cite{ning}.

\begin{figure}[t]
\includegraphics[scale=.45]{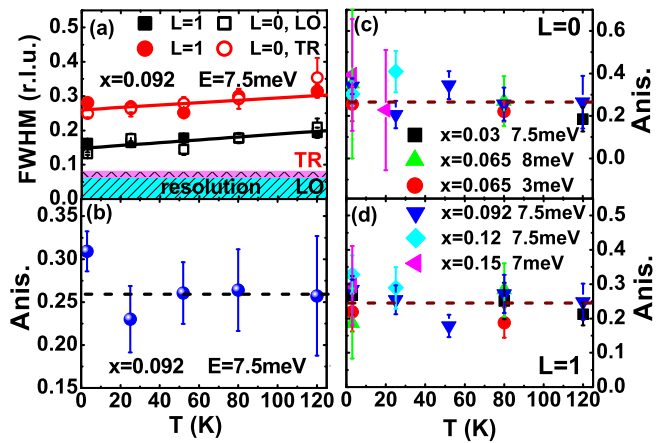}
\caption{ (Color online)
(a) and (b) Temperature dependence of FWHM for the transverse
and longitudinal wave vector scans and the in-plane anisotropy for
BaFe$_{2-x}$Ni$_x$As$_2$ ($x=0.092$) at $E=7.5$ meV.  (c) and (d)
Temperature dependence of in-plane anisotropy for all dopings at
$L= 0$ and $L= 1$.
}
\end{figure}

To determine if the widths of spin excitations are affected by temperature,
we show in Fig. 11(a) the temperature dependence of the FWHM of the
longitudinal and transverse spin excitations for
BaFe$_{1.908}$Ni$_{0.092}$As$_2$ at $E=7.5$ meV.  On warming from 3 K to 120 K,
the spin excitation widths are essentially unchanged, and increase only
slightly.  Figures 11(b)-11(d) show that the spin excitation
anisotropy ratio $A$ is also  temperature independent to within the errors of our measurements.
This is consistent with earlier neutron scattering results on BaFe$_{1.85}$Co$_{0.15}$As$_2$ \cite{jtpark10}.

\section{Discussion and Conclusions}

Since the discovery of the static AF order in the parent compounds
of iron-based superconductors \cite{cruz,jzhao,qhuang}, a central
question has been whether the static AF order and associated spin
excitations can be entirely described by Fermi surface nesting
between the hole pockets near $\Gamma$ point and electron pockets
$M$ points of the Brillouin zone \cite{iimazin08,jdong08}, or
requires local moments as in the case of copper oxide
superconductors \cite{khaule08,qmsi08,cfang08,ckxu08}.  From recent
neutron scattering experiments on spin waves in BaFe$_2$As$_2$
\cite{lharriger} and spin excitations in optimally electron doped
BaFe$_{1.9}$Ni$_{0.1}$As$_2$ \cite{msliu12}, we see that the effect
of electron-doping on BaFe$_2$As$_2$ modifies spin waves below 100
meV, and does not change high-energy spin excitations.  These
results suggest that spin excitations in BaFe$_{2-x}$Ni$_{x}$As$_2$
have both itinerant and localized components \cite{msliu12}.  Given
the general agreement on the evolution of the observed spin
excitation anisotropy and the RPA calculations based on rigid band
shift and itinerant electrons, one may assume
 that itinerant electrons and Fermi surface nesting only affect
low-energy spin excitations, and high-energy spin waves and
excitations arise mostly from local moments and electron
correlations. These results are consistent with the idea that the
transversely elongated spin excitations in BaFe$_{2-x}$Ni$_x$As$_2$
are mostly due to intraorbital, but interband, scattering processes
in cases without perfect nesting \cite{jzhang10}.

The experimental observation of a rapid reduction in the spin wave
anisotropy gap upon electron-doping in BaFe$_{2-x}$Ni$_x$As$_2$
confirms the earlier results that the dominate effect of electron
doping is to reduce $c$-axis spin exchange coupling and change
three-dimensional spin waves into quasi-two-dimensional spin
excitations \cite{lharriger2}. Based on our systematic measurements
of the transverse and longitudinal widths of spin excitations for
different Ni-doping levels $x$, we find that the intrinsic
excitation widths increase linearly with $x$.  This is consistent
with the RPA calculation assuming that the effect of Ni-doping is to
increase electron Fermi pocket size via rigid band shift.  Although
the RPA calculated spin susceptibility occupies a larger part of the
Brillouin zone than that of the experiments, their electron-doping
dependences are rather similar.  These results suggest that
itinerant electron picture and Fermi surface nesting can capture an
important part of the physics in these materials.  Similar RPA
calculations for much higher energy spin excitations ($E>100$ meV)
give results that disagree with our measurements, thus confirming
the notion that the high-energy spin excitations in
BaFe$_{2-x}$Ni$_x$As$_2$ may originate from the local moments
instead of Fermi surface nesting and itinerant electrons
\cite{msliu12}.

In conclusion, we use inelastic neutron scattering to demonstrate
the presence of anisotropic in-plane spin excitations at low
energies in electron doped BaFe$_{2-x}$Ni$_{x}$As$_{2}$. The
excitation widths in both the transverse and longitudinal directions
increase linearly with doping level $x$, and having slightly larger
slope in the transverse direction. In the overdoped side with
$x=0.15$, we find evidence for the low-energy transverse
incommensurate spin excitations consistent with the RPA calculation.
Therefore, the in-plane spin excitation anisotropy increases
slightly with doping. For samples near optimal superconductivity, a
neutron spin resonance appears below $T_c$.  However, the intensity
of spin excitations decreases with increasing doping for samples
beyond optimal superconductivity, and vanishes near the overdoped
border of the SC dome. Therefore, our data support the view that
low-energy spin excitations are controlled by Fermi surface nesting
and itinerant electrons, and are important for superconductivity of
iron pnictides.

\section{Acknowledgments}

We thank Jiangping Hu, Daoxing Yao and Qimiao Si for helpful
discussions. The work at the Institute of Physics, Chinese Academy
of Sciences, is supported by the MOST (973 project: 2012CB821400,
2011CBA00110, and 2010CB833102) and NSFC (No.11004233). The work at
UTK is supported by the U.S. National Science Foundation through
grant numbers NSF-DMR-1063866 and NSF-OISE-0968226. TAM acknowledges
the Center for Nanophase Materials Sciences, which is sponsored at
Oak Ridge National Laboratory by the Scientific User Facilities
Division, Office of Basic Energy Sciences, U.S. Department of
Energy.


\end{document}